\documentclass[preprint,showpacs,preprintnumbers,amsmath,amssymb]{revtex4}
\usepackage{epsfig}
\usepackage{graphicx}
\usepackage{dcolumn}
\usepackage{bm}

\newcommand{\ord}{{\cal O}}
\def\beq{\begin{equation}}
\def\eeq#1{\label{#1}\end{equation}}
\def\eeqn{\end{equation}}
\newcommand\iden{\leavevmode\hbox{\small1\normalsize\kern-.33em1}}

\newcommand{\sq}{\sqrt{2}}
\newcommand{\bea} {\begin{eqnarray}}
\newcommand{\eea} {\end{eqnarray}}

\newcommand{\Lg}{{\mathcal L}}
\newcommand{\rd}{\partial}

\newcommand{\Gm}{\Gamma}
\newcommand{\sbt}{s_{\beta}}
\newcommand{\cbt}{c_{\beta}}
\newcommand{\tbt}{t_{\beta}}
\newcommand{\sbcb}{s_{\beta} c_\beta}

\let\jnfont=\rm
\def\NPB#1,{{\jnfont Nucl.\ Phys.\ B }{\bf #1},}
\def\PLB#1,{{\jnfont Phys.\ Lett.\ B }{\bf #1},}
\def\EPJC#1,{{\jnfont Eur.\ Phys.\ Jour.\ C }{\bf #1},}
\def\PRD#1,{{\jnfont Phys.\ Rev.\ D }{\bf #1},}
\def\PRL#1,{{\jnfont Phys.\ Rev.\ Lett.\ }{\bf #1},}
\def\MPLA#1,{{\jnfont Mod.\ Phys.\ Lett.\ A }{\bf #1},}
\def\JPG#1,{{\jnfont J.\ Phys.\ G }{\bf #1},}
\def\CTP#1,{{\jnfont Commun.\ Theor.\ Phys.\ }{\bf #1},}
\def\JHEP#1,{{\jnfont JHEP \ }{\bf #1},}
\def\NPPS#1,{{\jnfont Nucl.\ Phys.\ Proc.\ Suppl.\ }{\bf #1},}
\def\CPC#1,{{\jnfont Computl.\ Phys.\ Commun.\ }{\bf #1},}

\begin{document}

\title{\ \\[10mm] Higgs-pair Production and Decay in Simplest Little Higgs Model}

\author{Xiao-Fang Han$^1$, Lei Wang$^2$$^{,*}$\footnotetext{* Corresponding author.
Email address: leiwang@itp.ac.cn (L. Wang)}, Jin Min Yang$^1$ }

\affiliation{$^1$ Key Laboratory of Frontiers in Theoretical Physics,
Institute of Theoretical Physics, Academia Sinica,
             Beijing 100190, China \\
$^2$ Department of Physics, Yantai University, Yantai 264005, China
\vspace*{1cm}}

\begin{abstract}
In the framework of the simplest little Higgs model (SLHM), we study
the production of a pair of neutral CP-even Higgs bosons at the LHC.
First, we examine the production rate and find that it can be
significantly larger than the SM prediction. Then we investigate the
decays of the Higgs-pair and find that for a low Higgs mass their
dominant decay mode is $hh\to\eta\eta\eta\eta$ ($\eta$ is a CP-odd
scalar) while $hh\to b\bar{b}\eta\eta$ and $hh\to \eta\eta WW$ may
also have sizable ratios. Finally, we comparatively study the rates
of $pp\to hh \to b\bar{b}\tau^+ \tau^-$, $pp\to hh \to
b\bar{b}\gamma\gamma$, and $pp\to hh \to WWWW$ in the SLHM and the
littlest Higgs models (LHT). We find that for a light Higgs,
compared with the SM predictions, all the three rates can be sizably
enhanced in the LHT but severely suppressed in the SLHM; while for an
intermediately heavy Higgs, both the LHT and SLHM can enhance sizably
the SM predictions.
\end{abstract}


\maketitle

\section{Introduction}
Little Higgs theory \cite{LH} has been proposed as an interesting
solution to the hierarchy problem. So far various
realizations of the little Higgs symmetry structure have been
proposed \cite{otherlh,lht,lst,sst}, which can be categorized
generally into two classes \cite{smoking}.
One class use the product group, represented by the littlest Higgs model
\cite{lst}, in which the SM $SU(2)_L$ gauge group is
from the diagonal breaking of two (or more) gauge groups.
The other class use the simple group, represented by
the simplest little Higgs model (SLHM) \cite{sst},
in which a single larger gauge group is broken down to
the SM $SU(2)_L$. Of course, different realizations give different
phenomenology, which will be tested at the LHC

Since these little Higgs models mainly alter the properties of the
Higgs boson and the top quark, hints of these models may be
unravelled from various Higgs boson and top quark processes
\cite{higgs-top-lht}. The Higgs-pair production at the LHC, albeit
with a small production rate, is rather important because it will
provide a way to probe the Higgs self-coupling $\lambda$. With the
designed luminosity, it is possible for the LHC to establish that
the SM Higgs boson has a non-zero self-coupling and the ratio
$\lambda/\lambda_{SM}$ can be restricted to a range of $0-3.7$ at
$95\%$ confidence level if its mass is between 150 GeV and 200 GeV
\cite{higgsself}. Such Higgs-pair production is sensitive to new
physics and has been studied in various new physics models
\cite{nphh}. In the littlest Higgs models without and with
T-parity, this process was studied in \cite{lsthh} and
\cite{lhthh}, respectively. In this work, we study this process in
the SLHM. We will first examine the Higg-pair production rate in
the SLHM and compare with the SM prediction. Then we study the
decays of the Higgs-pair. Finally, we study the rates of $pp\to hh
\to b\bar{b}\tau^+ \tau^-$ $(b\bar{b}\gamma\gamma)$ and $pp\to hh
\to WWWW$, comparing the prediction of the SLHM with the littlest
Higgs models.

This work is organized as follows. In Sec. II we recapitulate the
SLHM. In Sec. III we calculate the Higgs-pair production cross
section at the LHC. In Sec. IV, we study the decays of the Higgs-pair
and the rates of
$pp\to hh \to b\bar{b}\tau^+ \tau^-$
$(b\bar{b}\gamma\gamma)$ and $pp\to hh \to WWWW$.
Finally, we give our conclusion in Sec. V.

\section{Simplest little Higgs model}
The SLHM is based on $[SU(3) \times U(1)_X]^2$ global symmetry. The
gauge symmetry $SU(3) \times U(1)_X$ is broken down to the SM
electroweak gauge group by two copies of scalar fields $\Phi_1$ and
$\Phi_2$, which are triplets under the $SU(3)$ with aligned VEVs
$f_1$ and $f_2$. The uneaten five pseudo-Goldstone bosons can be
parameterized as

\beq
\Phi_{1}= e^{ i\; t_\beta \Theta } \left(\begin{array}{c} 0 \\
0 \\ f_1
\end{array}\right)\;,\;\;\;\;
\Phi_{2}= e^{- i\; t_\beta \Theta} \left(\begin{array}{c} 0 \\  0 \\
f_2
\end{array}\right)\;,
\label{paramet}
\end{equation}
where
\begin{equation}
   \Theta = \frac{1}{f} \left[
        \left( \begin{array}{cc}
        \begin{array}{cc} 0 & 0 \\ 0 & 0 \end{array}
            & H \\
        H^{\dagger} & 0 \end{array} \right)
        + \frac{\eta}{\sqrt{2}}
        \left( \begin{array}{ccr}
        1 & 0 & 0 \\
        0 & 1 & 0 \\
        0 & 0 & 1 \end{array} \right) \right],
\end{equation}
$f=\sqrt{f_1^2+f_2^2}$ and $t_\beta\equiv tan\beta= f_2 / f_1$.
Under the $SU(2)_L$ SM gauge group, $\eta$ is CP-odd singlet, while
$H$ transforms as a doublet and can be identified as the SM Higgs
doublet. The kinetic term in the non-linear sigma model is \beq
\label{eq:Lg:gauge0} \Lg_\Phi = \sum_{j=1,2}\left| \left(\rd_\mu + i
g A^a_\mu T^a - i \frac{g_x}{3} B^x_\mu \right) \Phi_j \right|^2,
\end{equation}
where $g_x =g\tan \theta_W/ \sqrt{1-\tan^2 \theta_W/3}$ with
$\theta_W$ being the electroweak mixing angle. As $\Phi_1$ and
$\Phi_2$ develop their VEVs, the new heavy gauge bosons $Z'$, $Y^0$,
and $W^{'\pm}$ get their masses proportional to $f$.

The gauged $SU(3)$ symmetry promotes the SM fermion doublets into
$SU(3)$ triplets. There are two possible gauge charge assignments
for the fermions: the 'universal' embedding and the 'anomaly-free'
embedding.  The first choice is not favored by the
electroweak precision data \cite{sst}, so we focus on the second
way of embedding. The quark Yukawa interactions for the third generation
and the first two generations can be written respectively as
\bea
{\cal L}_3 &=& i \lambda_1^t t_1^c \Phi_1^{\dagger} Q_3
  + i \lambda_2^t t_2^c \Phi_2^{\dagger} Q_3
  + i  \frac{\lambda_d^m}{\Lambda}  d_m^c \epsilon_{ijk}
      \Phi_1^i \Phi_2^j Q_3^k + h.c.\,, \label{simtopyukawa}\\
{\cal L}_{1,2} &=&  i \lambda_1^{d_n} d_{1n}^c Q_n^{T} \Phi_1
  + i \lambda_2^{d_n} d_{2n}^c Q^{T}_n \Phi_2
  + i \frac{\lambda_{u}^{mn}}{\Lambda} u_m^c \epsilon_{ijk} \Phi_1^{*i}
    \Phi_2^{*j} Q_n^k + h.c.,\label{simucyukawa}
\eea where $n=1,2$ are the first two generations indices;
$i,j,k=1,2,3$; $Q_3=\{ t_L, b_L, i T_L\}$ and $Q_n = \{ d_{nL}, -
u_{nL}, i D_{nL}\}$; $d_m^c$ runs over $(d^c, s^c, b^c, D^c, S^c)$;
$d^c_{1n}$ and $d^c_{2n}$ are linear combinations of $d^c$ and $D^c$
for $n=1$ and of $s^c$ and $S^c$ for $n=2$; $u^c_m$ runs over $(u^c,
c^c, t^c, T^c)$. For simplicity, we assume the quark flavor mixing
are small and neglect the mixing effects. From Eqs.
(\ref{simtopyukawa}) and (\ref{simucyukawa}), we can get the Higgs
boson interactions and the mass terms for the three generations of
quarks: \bea \label{tTmixing} {\cal L}_t &\simeq&-f \lambda_2^t
\left[ x_\lambda^t c_\beta t_1^c(-s_1t_L
   +c_1T_L)+s_\beta t_2^c (s_2 t_L+ c_2 T_L)\right]+h.c.,\,\\
   \label{dDmixing}
{\cal L}_{d_n} &\simeq&-f \lambda_2^{d_n} \left[ x_\lambda^{d_n} c_\beta d_1^c
  (s_1 d_{nL}+c_1 D_{nL})+s_\beta d_2^c (-s_2 d_{nL}+c_2 D_{nL})\right]+h.c.,\,\\
{\cal L}_{q} &\simeq&-\frac{\lambda_q}{\Lambda}f^2 s_\beta c_\beta s_3 q^c q_{_L}
          +h.c. \ (q=u,c,b)\,
\eea
where
\bea
x_\lambda^t\equiv {\lambda_1^t \over \lambda_2^t},\ \
x_\lambda^{d_n}\equiv {\lambda_1^{d_n} \over \lambda_2^{d_n}},\ \
s_{\beta}\equiv\frac{f_2}{\sqrt{f^2_1+f^2_2}},\ \
c_{\beta}\equiv\frac{f_1}{\sqrt{f^2_1+f^2_2}},\ \ \nonumber\\
s_1\equiv \sin {t_\beta (h+v)\over \sqrt{2}f},\ \ s_2\equiv
\sin{(h+v) \over \sqrt{2}t_\beta f},\ \ s_3\equiv
\sin{(h+v)(t_\beta^2+1)\over \sqrt{2}t_\beta f}, \eea with $h$ and
$v$ being the neutral Higgs boson field and its VEV, respectively.
The mass eigenstates are obtained by mixing the corresponding
interaction eigenstates, e.g., the mass eigenstates $(t_{mL},
T_{mL})$ and $(t_m^c, T_m^c)$ are  respectively the mixtures of
$(t_{L}, T_{L})$ and $(t^c, T^c)$. The diagonalization of the mass
matrix in Eqs.(\ref{tTmixing}) and (\ref{dDmixing}) was performed
numerically in our analysis, and the relevant couplings with Higgs
boson can also be obtained without resort to any expansion of $v/f$.
Hereafter we denote the mass eigenstates without the subscript '$m$'
for simplicity.

The Yukawa and gauge interactions break the global symmetry and then
provide a potential for the Higgs boson. However, the
Coleman-Weinberg potential alone is not sufficient since the
generated Higgs mass is too heavy and the new CP-odd scalar $\eta$
is massless. Therefore, one can introduce a tree-level $\mu$ term
which can partially cancel the Higgs mass \cite{sst,kmanhetaeta}:
\beq -\mu^2 (\Phi^\dagger_1 \Phi_2 + h.c.) = - 2 \mu^2 f^2 \sbt\cbt
\cos\left( \frac{\eta}{\sq \sbt\cbt f} \right)
 \cos \left(
 \frac{\sqrt{H^\dagger H}}{f \cbt\sbt}
\right).
\end{equation}
The scalar potential becomes
\beq \label{eq:VCW}
V = - m^2 H^\dagger H + \lambda (H^\dagger H)^2
 - \frac{1}{2} m_\eta^2 \eta^2 +\lambda' H^\dagger H \eta^2 + \cdots,
\end{equation}
where
\beq \label{eq:msq:lambda}
m^2 = m_0^2 - \frac{\mu^2}{\sbcb}, \quad
\lambda =\lambda_0 - \frac{\mu^2}{12\sbt^3 \cbt^3f^2}, \quad
\lambda' = - \frac{\mu^2}{4 f^2 \sbt^3 \cbt^3},
\end{equation}
with $m_0$ and $\lambda_0$ being respectively the one-loop
contributions to the Higgs boson mass and the quartic couplings from
the contributions of fermion loops and gauge boson loops \cite{sst}.
The Higgs VEV, the Higgs boson mass and the mass of $\eta$ are given
by \beq \label{eq:vsq:mH:meta} v^2 = \frac{ m^2}{\lambda} , \quad
m_h^2 = 2 m^2 , \quad m_\eta^2 = \frac{\mu^2}{\sbcb} \cos\left(
\frac{v}{\sqrt{2} f \sbcb} \right).
\end{equation}
The Coleman-Weinberg potential involves the following parameters:
\beq \label{para}
f,~ x_\lambda^t,~ t_\beta,~\mu,~m_\eta,~m_h,v.
\end{equation}
Due to the modification of the observed $W$ gauge boson mass,
$v$ is defined as \cite{kmanhetaeta}
\beq \label{eq:v} v \simeq v_0
\left[ 1+ \frac{v_0^2}{12 f^2}\frac{\tbt^4-\tbt^2+1}{\tbt^2} -
\frac{v_0^4}{180 f^4}\frac{\tbt^8-\tbt^6+\tbt^4-\tbt^2+1}{\tbt^4}
\right],
\end{equation}
where $v_0=246$ GeV is the SM Higgs VEV. Assuming that there are
no large direct contributions to the potential from physics at the
cutoff, we can determine other parameters in Eq. (\ref{para}) from
$f$, $t_\beta$ and $m_h$ with the definition of $v$ in Eq. (\ref{eq:v}).

\section{Higgs-pair production at LHC}
\begin{figure}[tb]
\begin{center}
 \epsfig{file=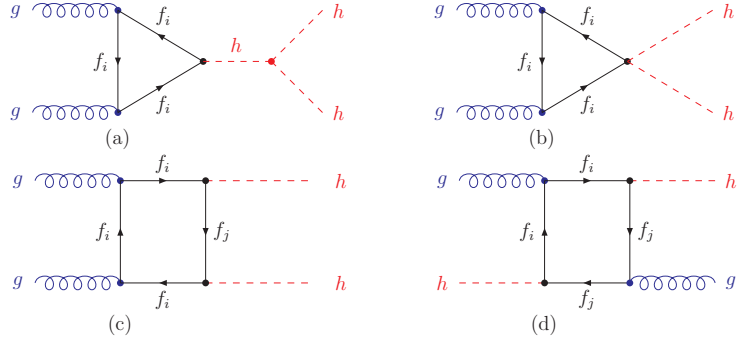,width=10cm}
\end{center}
\vspace{-1.0cm} \caption{Feynman diagrams for Higgs-pair production
via gluon-gluon fusion in the SLHM. Here $i,j=1,2$ with $(f_1,f_2)$
denoting $(t, T)$ or $(d, D)$ or $(s, S)$. The diagrams by
exchanging the two gluons or exchanging the two Higgs bosons in
(c,d) are not shown here.} 
\label{fmgg}
\end{figure}

\begin{figure}[tb]
\begin{center}
 \epsfig{file=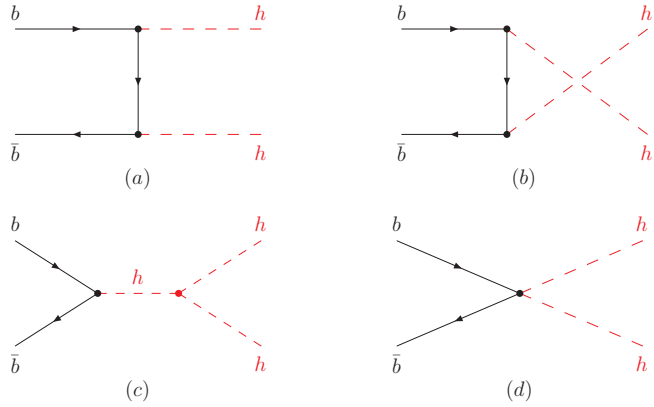,width=9cm}
\end{center}
\vspace{-1.0cm} \caption{Feynman diagrams for
Higgs-pair production via $b\bar{b}$ annihilation in the SLHM. }
\label{fmbb}
\end{figure}

At the LHC the Higgs-pair production can proceed through
gluon-gluon fusion and $b\bar{b}$ annihilation, as shown in Figs. (\ref{fmgg})
and (\ref{fmbb}), respectively. For the $b\bar{b}$
annihilation process, the SLHM can give the additional contributions
through the tree-level $hhb\bar b$ coupling and the modified $hb\bar
b$ coupling. For the gluon-gluon fusion process, the top-quark loops
give additional contributions through the tree-level $hht\bar t$
coupling and the modified $ht\bar t$ coupling. In addition to the
top-quark loops, the loops of the new heavy partner quarks $T$, $D$ and $S$
also come into play. Due to the large top quark Yukawa
coupling and the large parton distribution function of gluon at the LHC,
the contributions of the gluon-gluon fusion process can be dominant over
$b\bar{b}$ annihilation process.


The calculations of the loop diagrams in Fig. \ref{fmgg} are
straightforward. Each loop diagram is composed of some scalar loop
functions \cite{Hooft} which are calculated by using LoopTools
\cite{Hahn}. The calculations are tedious and the analytical
expressions are lengthy, which are not presented here. The hadronic
cross section at the LHC is obtained by convoluting the parton cross
section with the parton distribution functions. In our calculations
we use CTEQ6L \cite{cteq} to generate the parton distributions with
the renormalization scale $\mu_R $ and the factorization scale
$\mu_F$ chosen to be $\mu_R = \mu_F = 2m_{h}$ and use the two-loop
running coupling constant $\alpha_{s}$ with $\alpha_{s}(m_{Z})=0.118$.

The SM input parameters relevant in our study are taken as
$m_t=171.2$ GeV and $m_{Z}=91.1876$ GeV \cite{pdg}. The free SLHM
parameters are $f, t_\beta, m_h, x_\lambda^{d}(m_D)$ and
$x_\lambda^{s}(m_S)$. As shown above, the parameters
$x_\lambda^t,~\mu,~m_\eta$ can be determined by $f$, $t_\beta$,
$m_h$ and $v$. The small mass of the $d$ ($s$) quark requires one of
the couplings $\lambda^{d}_1$ and $\lambda^{d}_2$ ($\lambda^{s}_1$
and $\lambda^{s}_2$) to be very small, so there is almost no mixing
between the SM down-type quarks and their heavy partners. We assume
$\lambda^{d}_1$ ($\lambda^{s}_1$) is small, and take
$x_\lambda^{d}=1.1\times 10^{-4}$ and $x_\lambda^s=2.1\times10^{-3}$,
which can make the masses of $D$ and $S$ in the range of 1-2 TeV
with other parameters fixed as in our calculations. In fact,
our results show that the contributions from $d$ and $D$ ($s$ and
$S$) are small compared with the effects from $t$ and $T$. So,
different choices of $x_\lambda^{d}$ and $x_\lambda^{s}$ do not
have sizable effects on our results.

Electroweak precision data can give the strong constraints on the
scale $f$. The \cite{sst} shows that the LEP-II data requires $f>2$
TeV. In addition, the contributions to electroweak precision data
can be suppressed by large $t_\beta$. Ref. \cite{f4.5} gives a lower
bound of $f>4.5$ TeV from the oblique parameter $S$ while a recent
study of $Z$ leptonic decay gives a stronger bound of $f>5.6$ TeV
\cite{f5.6}. Considering the above bounds, we take $f=4$ TeV or
$f=5.6$ TeV with a large $\tan\beta$ to illustrate our results.

\begin{figure}[tb]
\begin{center}
 \epsfig{file=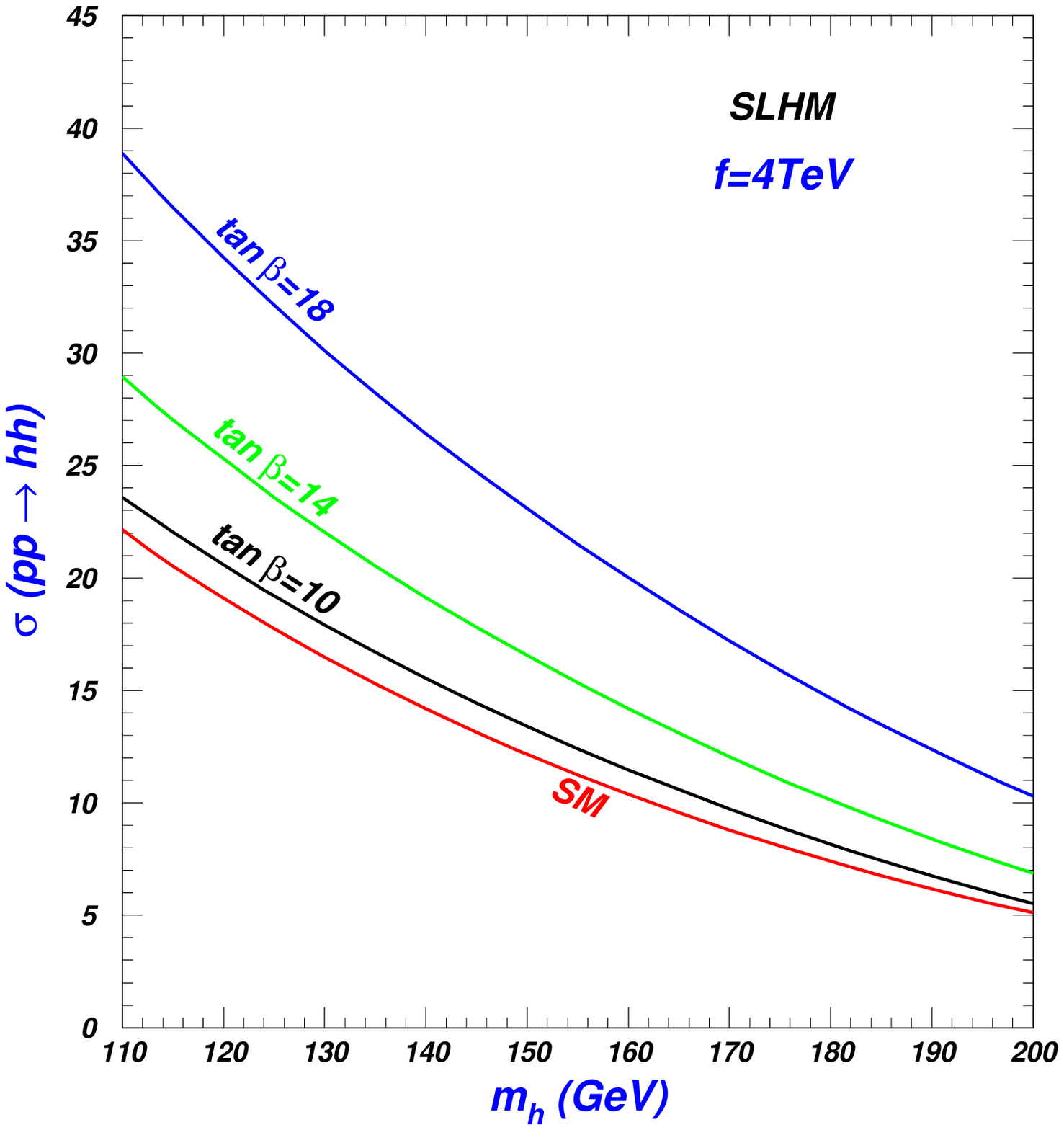,width=7cm}
 \epsfig{file=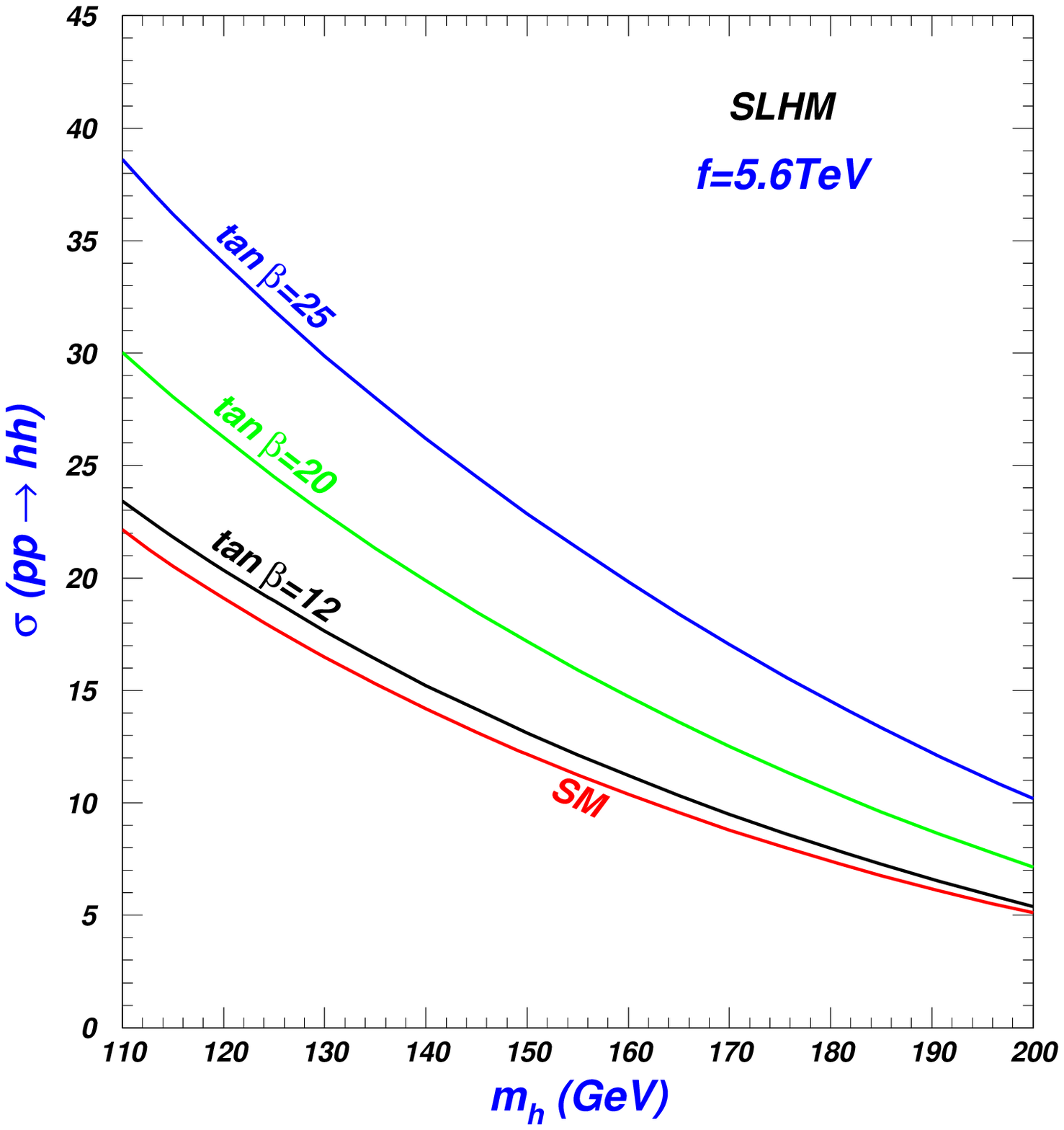,width=7cm}
\end{center}
\vspace{-1.0cm} \caption{Hadronic cross section of Higgs-pair
production at the LHC versus the Higgs boson mass.} \label{cross}
\end{figure}

In Fig. \ref{cross}, we take several values of $\tan\beta$ and plot
the hadronic cross section of Higgs-pair production at the LHC
versus the Higgs boson mass. We find that compared with the SM
prediction, the cross section in the SLHM can be significantly
enhanced for a large $\tan\beta$. For example, with $\tan\beta$=18
(25) and $f=4$ TeV (5.6 TeV), the cross section can be enhanced by
$80\%$ for $m_h=110$ GeV. Of course, for the perturbation to be
valid, $\tan\beta$ cannot be too large for fixed $f$. As shown in
Eq. (\ref{eq:v}), the correction to the Higgs VEV is proportional to
$\tan^2\beta v_0^2/f^2$. If we require
$\ord(v_0^4/f^4)/\ord(v_0^2/f^2) < 0.1$ in the expansion of $v$, the
value of $\tan\beta$ should be below 20 (28) for $f=4$ TeV (5.6
TeV). For a larger $f$, the value of $\tan\beta$ can be larger and
cancel partially the suppression of $v/f$. Therefore, the maximal
value of the cross section does not always decrease with increasing
of $f$.

\section{Final states of Higgs-pair production}
The Higgs-pair production can give various final states, depending on
the decay modes of the Higgs boson.
The SLHM corrections to the tree-level decays $h\to f
\bar{f},WW,ZZ$ are mainly from the corresponding modified couplings:
\beq
\Gamma(h \to XX)= \Gamma(h \to XX)_{SM}(g_{hXX}/g_{hXX}^{SM})^2,
\end{equation}
where $XX$ denotes $WW$, $ZZ$ or fermion pairs, and $\Gamma(h \to
XX)_{SM}$ is the SM decay width. $g_{hXX}$ and $g_{hXX}^{SM}$ are
the couplings of $hXX$ in the SLHM and SM, respectively. The
couplings $g_{hWW}$ and $g_{hZZ}$ can be found in
\cite{kmanhetaeta}.

For the low Higgs mass, the loop-induced decay $h \to gg$ will be
also important. In addition to the top quark loops, the loops of new
heavy quarks ($T$, $D$, $S$) come into play.
For another important loop-induced decay mode $h \to \gamma\gamma$,
in addition to the contributions of top quark and $W$ boson, the new
charged heavy fermions ($T$, $D$, $S$) and gauge bosons $W'^{\pm}$
will make contributions. Following the approach in \cite{hrr}, the partial
decay width of  $h \to \gamma\gamma$ can be calculated at one-loop level.
For the SM decay channels, the relevant higher order QCD and electroweak
corrections are considered using the code Hdecay \cite{hdecay}.

In addition to the SM decay modes, the Higgs boson in the SLHM
has two new important decay modes, $h\to \eta\eta$ and $h \to Z\eta$,
in the kinematically allowed parameter space. Their partial widths
are given by
\bea \label{eq:Gamma:new}
\Gm(h \to \eta\eta) &=& \frac{{\lambda'}^2}{8\pi}\frac{v^2}{m_h} \sqrt{1-x_\eta},\nonumber\\
\Gamma( h \to Z \eta) &=& \frac{m_h^3}{32 \pi f^2}
  \left( t_\beta - \frac{1}{t_\beta} \right)^2 \,
  \lambda^{3/2} \left(1, \frac{m_Z^2}{m_h^2}, \frac{m_\eta^2}{m_h^2}
 \right ),
\eea
where $x_\eta =4m_\eta^2/m_h^2$ and $\lambda (1,x,y) = (1-x-y)^2 - 4 xy$.
These two decay channels can be dominant in the allowed parameter space
\cite{kmanhetaeta} and provide some new signatures of Higgs-pair production.

\begin{figure}[tb]
\begin{center}
 \epsfig{file=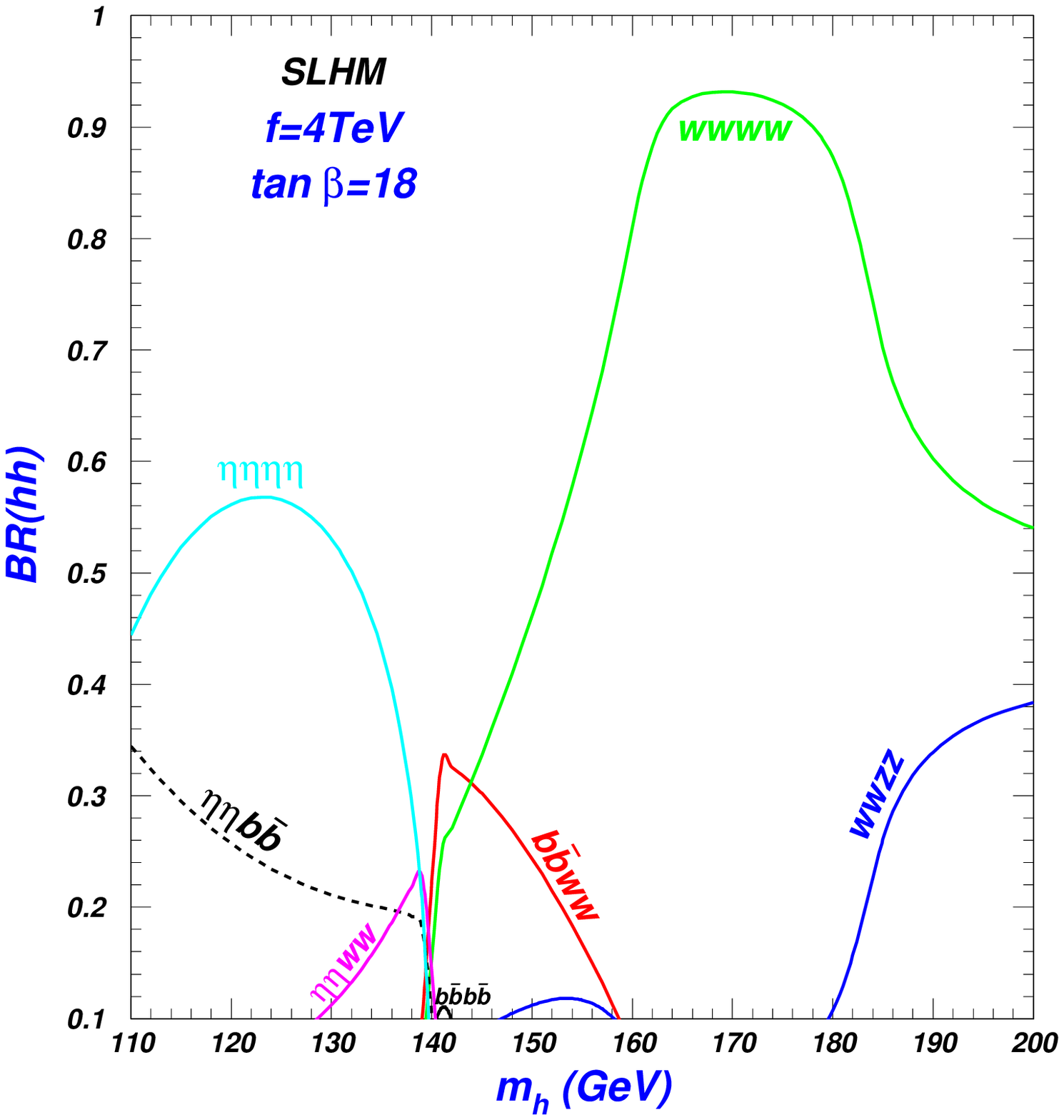,width=7cm}
 \epsfig{file=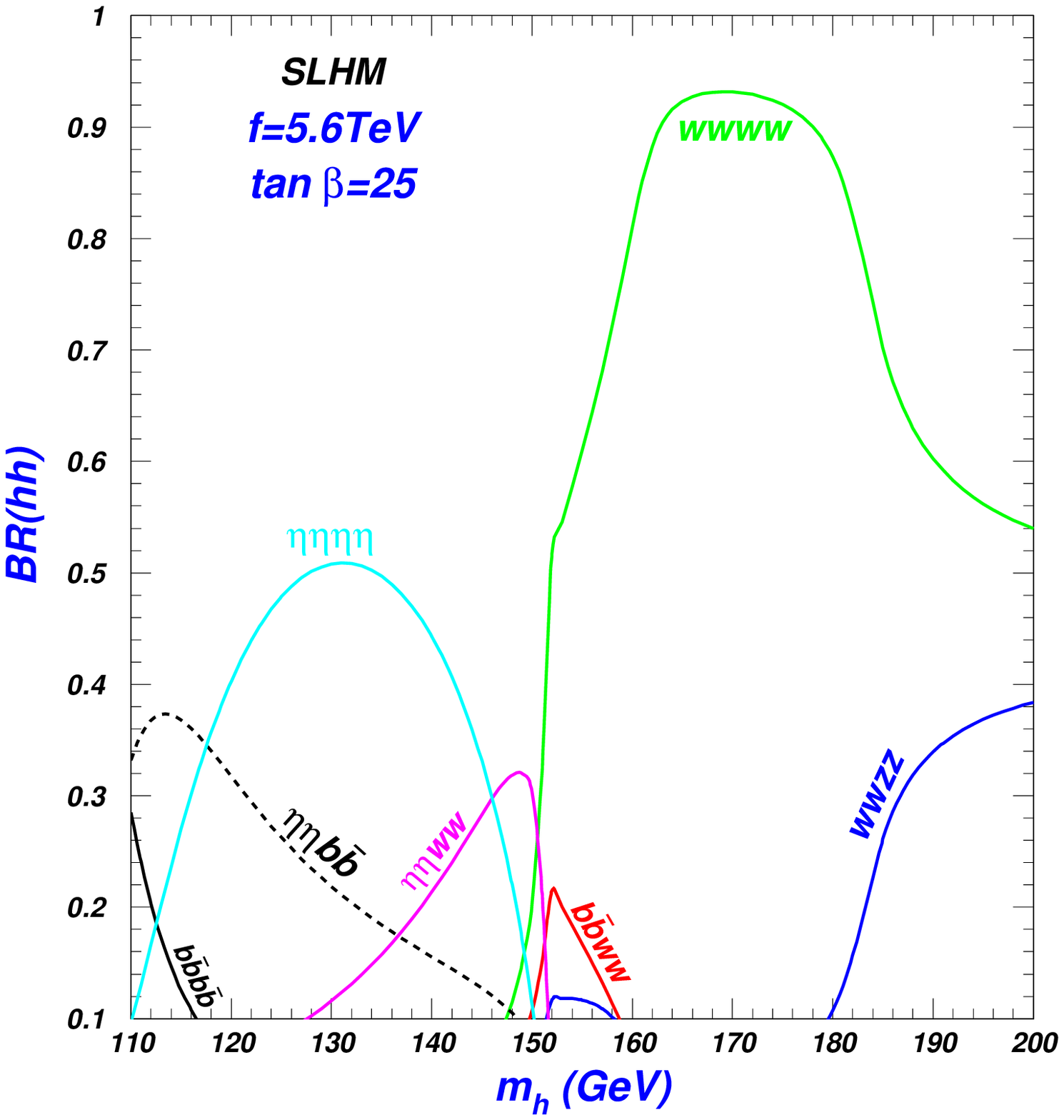,width=7cm}
\end{center}
\vspace{-1.0cm} \caption{The decay branching ratios of Higgs-pair as a function
of  the Higgs boson mass. } \label{decay}
\end{figure}

Fig. \ref{decay} shows the decay branching ratios of Higgs-pair versus the
Higgs boson mass (we only plot the decay modes with branching ratio above 0.1).
We see that the dominant decay channel is $hh\to WWWW$ for 150 GeV $<m_h<$ 200 GeV,
similar to the SM prediction; but for Higgs mass below $140$ GeV, the decay
$hh\to \eta\eta\eta\eta$ will dominate over $hh\to b\bar{b}b\bar{b}$
which has the largest branching ratios in the SM.
Besides, the decays $hh\to b\bar{b}\eta\eta$ and $hh\to \eta\eta WW$ can also
be important, whose branching ratios can be much larger than the
decay $hh\to b\bar{b}b\bar{b}$ in some part of the parameter space.

The decays of $\eta$ have been studied in \cite{wk,kmanetadecay}.
For 10 GeV $<m_\eta<$ 100 GeV, $\eta$ decays mainly into $b\bar{b}$,
$\tau^+ \tau^-$ or $gg$. The branching ratio of $\eta\to \tau^+
\tau^-$ is about $10\%$ of $\eta\to b\bar{b}$. With increasing of
$m_\eta$, the branching ratios of $\eta\to b\bar{b}$ and $\eta\to
\tau^+ \tau^-$ decrease while the decay $\eta \to gg$ increases and
may surpass the ratio of $\eta\to b\bar{b}$.

\begin{figure}[tb]
\begin{center}
 \epsfig{file=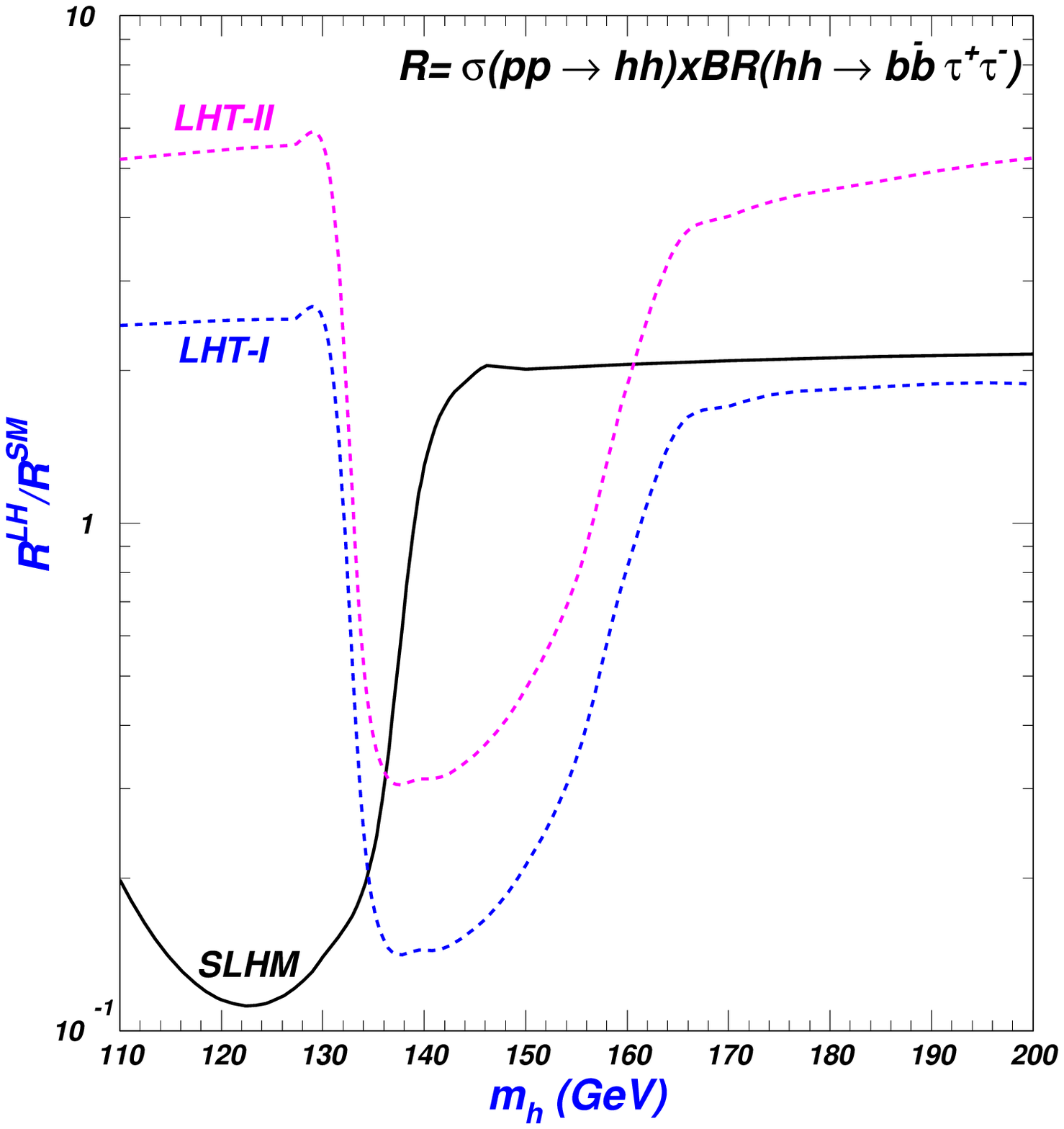,width=5.3cm}
 \epsfig{file=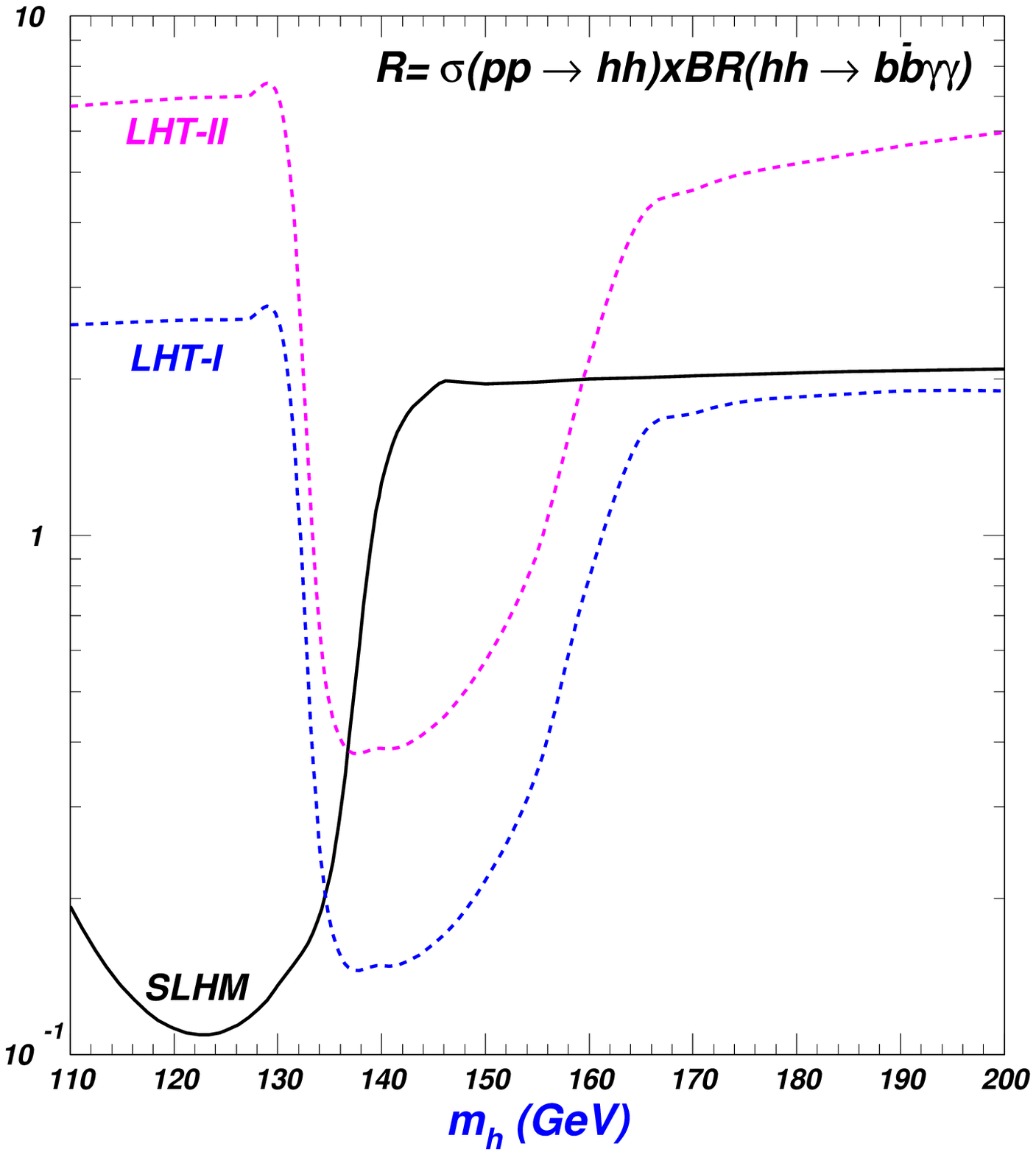,width=5.05cm}
 \epsfig{file=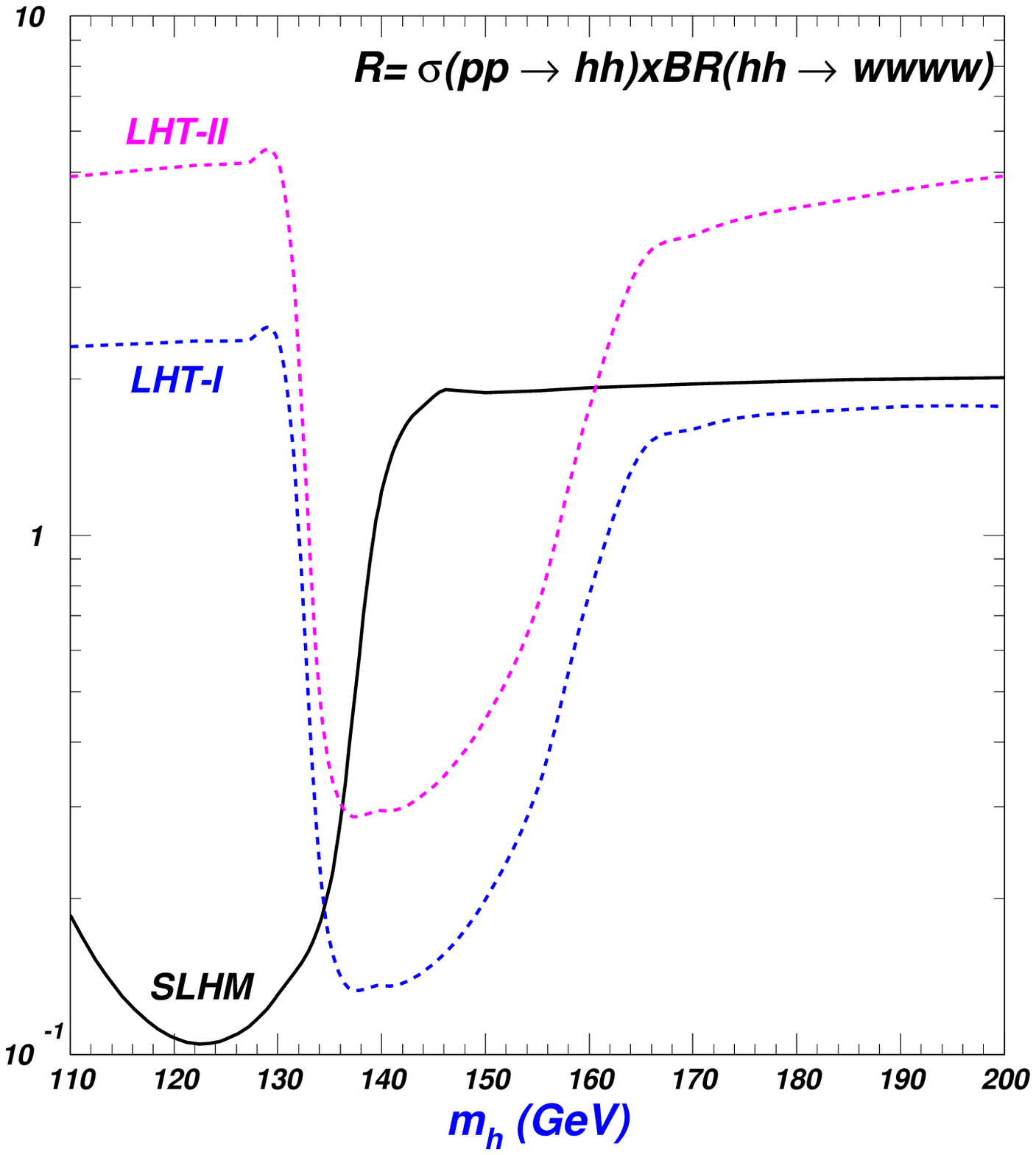,width=5.05cm}
\end{center}
\vspace{-1.0cm} \caption{The rates of $\sigma(pp\to hh)\times
BR(hh\to b\bar{b}\tau^+ \tau^-)$, $\sigma(pp\to hh)\times BR(hh\to
b\bar{b}\gamma\gamma)$ and $\sigma(pp\to hh)\times BR(hh\to WWWW)$
 in LHT-I, LHT-II and SLHM, normalized to the SM prediction. Here,
$\tan\beta$ is fixed as 18 and
 $f$ is taken as its lower bound, which is 4 TeV for the SLHM
and 500 GeV for LHT-I and LHT-II \cite{flht-i}.}
\label{crossbr}
\end{figure}

In the SM the promising channels are $pp\to hh \to b\bar{b}\tau^+
\tau^-$ $(b\bar{b}\gamma\gamma)$ for $m_h<140$ GeV \cite{bbtautau}
and $pp\to hh \to WWWW$ for $150$ GeV $<m_h<200$ GeV
\cite{higgsself}. In Fig. \ref{crossbr} we plot the rates of
$\sigma(pp\to hh)\times BR(hh\to b\bar{b}\tau^+ \tau^-)$,
$\sigma(pp\to hh)\times BR(hh\to b\bar{b}\gamma\gamma)$ and
$\sigma(pp\to hh)\times BR(hh\to WWWW)$ normalized to the SM
predictions, and compare the SLHM results with the predictions of
two types of littlest Higgs models with T-parity (LHT-I and LHT-II).
The detailed descriptions of LHT-I and LHT-II can be found in
\cite{lht,lhtintro}. We see that for $m_h<130$ GeV all the three
rates can be enhanced sizably in LHT-I and LHT-II, but suppressed
significantly in the SLHM. For the larger value of $m_h$, both the
SLHM and LHT-I/LHT-II can enhance sizably the SM predictions (in the
SLHM for $m_h>150$ GeV, while in the LHT for $m_h>170$ GeV).

\section{Conclusion}
In the framework of the simplest little Higgs model (SLHM), we
studied the production of a pair of neutral CP-even Higgs bosons at
the LHC and obtained the following observations: (i) The Higgs-pair
production rate in the SLHM can be significantly larger than the SM
prediction; (ii) For a low Higgs mass the dominant decay mode of
Higgs-pair is $hh\to\eta\eta\eta\eta$ ($\eta$ is a CP-odd scalar)
while $hh\to b\bar{b}\eta\eta$ and $hh\to \eta\eta WW$ may also have
sizable ratios; (iii) For a light Higgs boson all the rates of
$pp\to hh \to b\bar{b}\tau^+ \tau^-$, $pp\to hh \to
b\bar{b}\gamma\gamma$ and $pp\to hh \to WWWW$ can be sizably
enhanced in the littlest Higgs models but severely suppressed in the
SLHM; while for an intermediately heavy Higgs, all the three rates can be
sizably enhanced in the littlest Higgs models and the SLHM.

\section*{Acknowledgment}
We thank C. P. Yuan for discussions.
This work was supported in part by the National Natural
Science Foundation of China under grant Nos. 10821504, 10725526 and 10635030.


\begin{thebibliography}{99}

\bibitem{LH}
N.~Arkani-Hamed, A.~G.~Cohen and H.~Georgi,
  Phys.\ Lett.\ B {\bf 513}, 232 (2001);
N.~Arkani-Hamed, A.~G.~Cohen, E.~Katz, A.~E.~Nelson, T.~Gregoire and
J.~G.~Wacker,
  JHEP {\bf 0208}, 021 (2002).

\bibitem{otherlh}
  D.~E.~Kaplan and M.~Schmaltz,
  JHEP {\bf 0310},  039 (2003);
  I.~Low, W.~Skiba, and D.~Smith,
  Phys.\ Rev.\ D {\bf 66}, 072001 (2002);
  S.~Chang and J.~G.~Wacker,
  Phys.\ Rev.\  D {\bf 69}, 035002 (2004);
  T.~Gregoire, D.~R.~Smith, and J.~G.~Wacker,
  Phys.\ Rev.\ D {\bf 69}, 115008 (2004);
  W.~Skiba and J.~Terning,
  Phys.\ Rev.\ D {\bf 68}, 075001 (2003);
  S.~Chang,
  JHEP {\bf 0312}, 057 (2003);
  H. Cai, H.-C. Cheng, and J. Terning, \JHEP0905, 045 (2009);
  A. Freitas, P. Schwaller, and D. Wyler, arXiv:0906.1816.

\bibitem{lht} H. C. Cheng, I. Low, \JHEP0309, 051 (2003); JHEP 0408, 061 (2004);
              H. C. Cheng, I. Low and L. T. Wang, \PRD74, 055001 (2006).

\bibitem{lst} N.~Arkani-Hamed, A.~G.~Cohen, E.~Katz and A.~E.~Nelson,
  JHEP {\bf 0207}, 034 (2002).

\bibitem{sst}
M.~Schmaltz,
  JHEP {\bf 0408}, 056 (2004).

\bibitem{smoking} T.~Han, H.~E.~Logan and L.~T.~Wang,
  JHEP {\bf 0601}, 099 (2006).

\bibitem{higgs-top-lht} See, e.g.,
               C. R. Chen, K. Tobe, C. P. Yuan, \PLB640, 263 (2006);
               K. Hsieh, C. P. Yuan, \PRD78,053006 (2008);
               C. O. Dib, R. Rosenfeld, A. Zerwekh, \JHEP0605, 074 (2006);
               L. Wang, {\it et al.}, \PRD75, 074006 (2007);  \PRD79, 055013 (2009);
               X. F. Han, L. Wang, J. M. Yang, \PRD78, 075017 (2008); arXiv:0903.5491;
               R. S. Hundi, B. Mukhopadhyaya, A. Nyffeler, \PLB649, 280 (2007);
               X. Wang, Y. Zhang, H. Jin, Y. Xi, \NPB810, 226 (2009); \NPB807, 210 (2009);
               C.-X. Yue, H.-D. Yang, W. Ma, \NPB818, 1 (2009);
               H. S. Hou, \PRD75, 094010 (2007);
               P. kai, et al., \PRD76, 015012 (2007).

\bibitem{higgsself} U. Baur, T. plehn, and D. Rainwater, \PRL89, 151801 (2002); \PRD67, 033003 (2003).

\bibitem{nphh}
      T. Plehn, M. Spira, and P. M. Zerwas, \NPB479, 46 (1996);
      J. Yi et al., \JPG 23, 385 (1997); \JPG 23, 1151 (1997);
      S. Dawson, S. Dittmaier, and M. Spira, \PRD58, 40 (1998);
      E. W. N. Glover and J. J.  van der Bij, \NPB309, 282 (1988);
      A. Krause, T. Plehn, M. Spira, and P. M. Zerwas, \NPB519, 85 (1998);
      A. A. B. Bendezu and B. A. Kniehl, \PRD64, 035006 (2001);
      W. Ma, C.-X. Yue, and Y.-Z. Wang, \PRD79, 095010 (2009);
      H. de Sandes and R. Rosenfeld, \PLB659, 323 (2008);
      A. Arhrib, et al., JHEP {\bf 0908}, 035 (2009);
      S. Kanemura and K. Tsumura, \EPJC63, 11 (2009).

\bibitem{lsthh} J. J. Liu, {\it et al.}, \PRD70, 015001 (2004).

\bibitem{lhthh} L. Wang,  {\it et al.}, \PRD76, 017702 (2007); \PRD77, 015020 (2008);

\bibitem{kmanhetaeta} K. Cheung and J. Song, \PRD76, 035007 (2007).

\bibitem{Hooft} G.~'t Hooft and M.~J.~G.~Veltman, \NPB153, 365 (1979).

\bibitem{Hahn} T.~Hahn and M.~Perez-Victoria,
               Comput.\ Phys.\ Commun.\  {\bf 118}, 153 (1999);
               T.~Hahn,  Nucl.\ Phys.\ Proc.\ Suppl.\  {\bf 135}, 333 (2004).

\bibitem{cteq} J.~Pumplin {\it et al.}, JHEP {\bf 0602}, 032 (2006).

\bibitem{pdg} C. Amsler {\it et al.}, \PLB667, 1 (2008).

\bibitem{f4.5} G. Marandella, C. Schappacher and A. Strumia, \PRD72, 035014 (2005).

\bibitem{f5.6} A. G. Dias, C. A. de S. Pires, P. S. Rodrigues da Silva, \PRD77, 055001 (2008).

\bibitem{hrr} T. Han, H. E. Logan, B. McElrath and L.-T. Wang, \PLB563, 191 (2003);
              Erratum-ibid. B {\bf603}, 257 (2004).

\bibitem{hdecay} A. Djouadj, J. Kalinowski and M. Spira, \CPC108, 56 (2006).

\bibitem{flht-i} J. Hubisz, P. Meade, A. Noble, M. Perelstein, \JHEP0601, 135 (2006).

\bibitem{wk} W. Kilian, D. Rainwater and J. Reuter, \PRD71, 015008 (2005).

\bibitem{kmanetadecay} K. Cheung, J. Song, P. Tseng and Q.-S. Yan, \PRD78, 055015 (2008).

\bibitem{bbtautau} U. Baur, T. Plehn and D. L. Rainwater, \PRD68, 033001 (2003); \PRD69, 053004 (2004).

\bibitem{lhtintro} See, e.g., J. Hubisz and P. Meade, \PRD71, 035016 (2005);
                   M. Blanke, \JHEP0701, 066 (2007).

\end{thebibliography}
\end{document}